\documentstyle[epsf,12pt]{article}
\def\beq{\begin{equation}}
\def\eeq{\end{equation}}
\def\bea{\begin{eqnarray}}
\def\eea{\end{eqnarray}}
\def\bq{\begin{quote}}
\def\eq{\end{quote}}
\catcode`\@=11
\@addtoreset{equation}{section}
\def\theequation{\arabic{section}.\arabic{equation}}

\begin{document}
\renewcommand{\thefootnote}{\dag}
\pagestyle{empty}
\vspace*{-1.5in}
\begin{flushright}
{CERN-TH-95-84}\\
{SHEP 95-14}
\end{flushright}
\begin{center}
{\bf SEARCH FOR LARGE RAPIDITY GAP EVENTS IN ${\bf e^+e^-}$ ANNIHILATION}\\
\vspace*{1cm} {\bf John Ellis} \\
\vspace*{0.3cm}
{\it Theoretical Physics Division, CERN} \\
{\it CH - 1211 Geneva 23} \\
and \\
\vspace*{0.3cm}
{\bf Douglas A. Ross}\\
{\it Physics Department, University of Southampton} \\
{\it Highfield, Southampton SO17 1BJ, England} \\
\vspace*{1cm}
{\bf ABSTRACT} \\ \end{center}
We investigate the cross-section for the production of a low-mass
colour-singlet cluster in $e^+e^-$ annihilation with a large
rapidity gap between the colour-singlet cluster and the other jets.
It is argued that such events are the cross-channel analogue of
large-rapidity-gap events in deep-inelastic scattering, and therefore
could in principle be used to investigate the analytic
continuation of the BFKL pomeron
to the positive-$t$ kinematic regime, where one would expect
the trajectory to pass through glueball states. The cross section
can be calculated in perturbative QCD, so that the infrared scale
arising from non-perturbative effects, which prevents an
exponential fall-off
with rapidity gap in the case of deep-inelastic scattering, is absent
in $e^+ e^-$ annihilation. Correspondingly, the cross section for such
 events decreases  rapidly with increasing rapidity gap.
\noindent

\vspace*{3cm}

\begin{flushleft}
CERN-TH-95-84 \\
{SHEP 95-14}\\
4 April 1995
\end{flushleft}
\vfill\eject

\setcounter{page}{1}
\pagestyle{plain}
 {\newcommand{\la}{\mbox{\raisebox{-.6ex}{$\stackrel{<}{\sim}$}}}}
{\newcommand{\ga}{\mbox{\raisebox{-.6ex}{$\stackrel{>}{\sim}$}}}}
\voffset -1in
\textheight=20cm
\vskip2.0cm

\section{Introduction}

The Pomeron is currently experiencing a renaissance, both theoretical
- via several different perturbative approaches - and experimental
- stimulated by observations at HERA. In the theoretical domain we
mention in particular the BFKL Pomeron \cite{BFKL} obtained by summing
perturbative logarithms in longitudinal momenta, $x$, and attempts to
interpolate between  the (BFKL) integral equation which describes
 this Pomeron and the GLAP evolution equation \cite{GLAP} which sums
perturbative logarithms in transverse momenta, $k_\perp$. In the
experimental domain we mention in particular the  growth
of the structure function at low values of Bjorken $x$ and increasing
$Q^2$, which may  be a signal for the BFKL Pomeron, and the
observation of deep inelastic final states with large rapidity gaps,
which likewise may be revealing point-like structure within the Pomeron.
\footnote{We emphasize here that the structure functions at low $x$ are
related by the optical theorem to the imaginary part of a single Pomeron
 exchange
amplitude, whereas in the case of events with large rapidity-gaps it
is the {\it amplitude} that is dominated by Pomeron exchange.}

The latter observations were preceded by the discovery of large $p_\perp$
hadronic jets in large rapidity-gap events at the SPS $p \bar{p}$
collider, which could be interpreted in terms of a hard point-like
Pomeron structure function. These were followed by the observation
of large rapidity-gaps in events with large $p_\perp$ hadronic jets at the
FNAL $p \bar{p}$ collider. In view of the existence of large rapidity-gap
events in $ep$ and $p \bar{p}$ collisions, it is natural to ask whether
analogous events occur in high energy $e^+ e^-$ collisions. This question
has been asked theoretically by Bjorken, Brodsky and Lu \cite{BBL},
who predicted a very small rate for large rapidity-gap events in
Z decays, and indeed none have been reported by any experiment.

The purpose of this paper is to re-examine theoretically the possible 
existence
 of
large rapidity-gap events in $e^+ e^-$ annihilation, building on the 
increased
insight into the Pomeron provided by recent theoretical studies and HERA
 measurements.
Large rapidity-gap events in $e^+ e^-$ annihilation would involve the 
production
of an isolated cluster of hadrons that would in some sense constitute the
direct channel ($m^2>0$) analogue of the crossed channel ($t<0$) Pomeron.
Indeed if the cluster were a single particle this would probably be a 
glueball
state which lies on the Pomeron trajectory for positive $t$. The relation
between these time-like and space-like regions may cast new light on
both perturbative and non-perturbative aspects of the Pomeron.

\begin{figure}
\begin{center}
\leavevmode
\hbox{\epsfxsize=1.8 in
\epsfbox{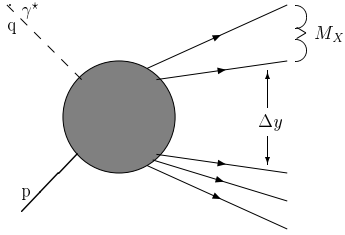}}
\end{center}
\caption{Rapidity-gap events in $ep$ scattering.} \label{fig1} \end{figure}

The basic mechanism of the HERA large rapidity-gap
events is illustrated in Fig.\ref{fig1}.  A photon of high virtuality
$Q \equiv \sqrt{-q^2}$ strikes a proton, producing a hadronic system
of large mass $W=\sqrt{(q+p)^2}$, that comprises of two components - an
 undetected
proton fragment separated by a large rapidity gap ($\Delta y$) from a
hadronic cluster of mass, $M_X$. Such final states appear to
constitute a finite fraction of the  total deep-inelastic structure
function, $F_2$ in the double scaling limit, $W/Q=1/x-1 \rightarrow \infty$,
$M_X/Q=1/\beta-1$ fixed, and $Q \rightarrow \infty$.  In Regge lore
$$ F_2(x,Q) \propto \left( \frac{1}{x} \right)^{\alpha_P(0)-1} $$
at small $x$ and $Q$ fixed, where $\alpha_P(t)$ is the Pomeron trajectory 
which
 is
treated here as a simple Regge pole. The contribution of large rapidity gap
 events
would be parametrized in this framework as
$$ F_2^{LRG}(x,Q) \propto \left( \frac{1}{x} \right)^{2\alpha_P(t)-1},$$
where $t$ is the momentum transferred between the initial proton and its 
final
 state fragment.
Since $|t|$ is small in the bulk of these events, the approximate
 $x$-independence of the fraction
$F_2^{LRG}/F_2$ corresponds to $\alpha_P(0) \approx 1$ as inferred from 
hadronic
phenomenology \cite{LD}. Thus the probability for these large rapidity gap
 events is
proportional to $\exp ( - c \Delta y)$ with $c=1-\alpha_P(0) \approx 0$, 
so we
 do not get
a substantial decrease in the number of rapidity-gap events as the rapidity 
gap
is increased. If Pomeron exchange factorizes in the cross channel, which is
 consistent with hadron
phenomenology \cite{LD} (but is {\it not} yet confirmed in deep inelastic
 scattering and which
furthermore cannot be understood within the context of perturbative QCD), 
then
 the structure
function of the Pomeron, $F_2^P(\beta,Q)$, may be extracted from the
 large-rapidity-gap events
at HERA.

\begin{figure}
\begin{center}
\leavevmode
\hbox{\epsfxsize=1.2 in
\epsfbox{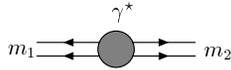}}
\end{center}
\caption{Events with two low-mass clusters in $e^+ e^-$ annihilation
 (viewed in the centre-of-mass frame).}
\label{fig2} \end{figure}

Bjorken, Brodsky and Lu \cite{BBL} considered a kinematical configuration in
 which
$e^+ e^-$ goes to two low-mass clusters, as illustrated in Fig.\ref{fig2}.
\begin{figure}
\begin{center}
\leavevmode
\hbox{\epsfxsize=1.5 in
\epsfbox{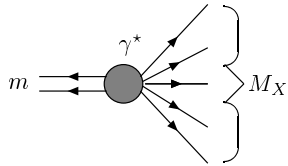}}
\end{center}
\caption{Large-rapidity-gap configuration in $e^+e^-$ annihilation 
(again viewed
 in the
centre-of-mass frame).} \label{fig3} \end{figure}
 We believe that a closer
analogue in $e^+ e^-$ annihilation to the HERA large rapidity-gap 
observations
 (in which $Q$ and
$M_X$ are both large, with $t$ fixed) is provided by the kinematical
 configuration illustrated in
Fig. \ref{fig3}, where only one of the two produced clusters is required 
to have
 a small mass ($m$), while the
other is allowed to have a large mass, $M_X \sim {\cal O}(Q)$. The small-mass
 cluster looks like
a colour-singlet jet with scaled energy fraction, $x_3=1-M_X^2/Q^2$.
\begin{figure}
\begin{center}
\leavevmode
\hbox{\epsfxsize=.25 in
\epsfbox{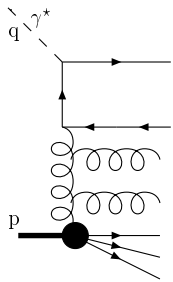}}
\end{center}
\caption{Photon-gluon fusion.} \label{fig4} \end{figure}
In QCD perturbation theory, the  
deep-inelastic structure function at low $x$ is
 dominated by photon-gluon
 fusion as illustrated in Fig. \ref{fig4}.
\begin{figure}
\begin{center}
\leavevmode
\hbox{\epsfxsize=2.8 in
\epsfbox{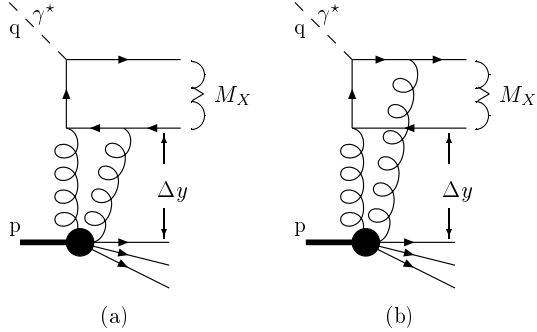}}
\end{center}
\caption{Leading-order contribution to large-rapidity-gap events in
 deep-inelastic scattering.}
\label{fig5} \end{figure}
The lowest-order perturbative contribution to large rapidity-gap
deep-inelastic events involves two gluon exchanges as illustrated in
 Fig.\ref{fig5}.
The standard three-jet final state of Fig. \ref{fig6} is the closest $e^+ e^-$
analogue of the photon-gluon fusion diagram of Fig.\ref{fig4}.
\begin{figure}
\begin{center}
\leavevmode
\hbox{\epsfxsize=.16 in
\epsfbox{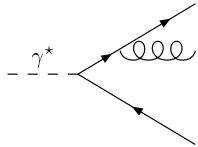}}
\end{center}  \vspace*{.3in}
\caption{ Standard three-jet events in $e^+ e^-$ annihilation.}
\label{fig6} \end{figure}
\noindent The lowest-order perturbative
contributions to the colour-singlet cluster
 production
in $e^+ e^-$ annihilation are those shown in Fig. \ref{fig7}, where
 Fig.\ref{fig7}(a),
in which a  ``glueball'  is produced, is very similar to the deep-inelastic
 two-gluon exchange
diagram of Fig. \ref{fig5}.
\begin{figure}
\begin{center}
\leavevmode
\hbox{\epsfxsize=3.6 in
\epsfbox{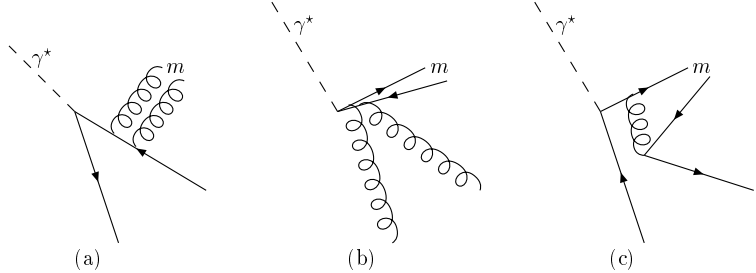}}
\end{center}
\caption{Single low-mass cluster events in $e^+ e^-$ annihilation.
For each type of event we display only one of the possible Feynman graphs.}
\label{fig7} \end{figure}

Here we make a few pertinent comments about the leading perturbative
 contribution to
deep-inelastic large rapidity-gap events due to the diagrams
 shown in Fig.\ref{fig5}.
If one calculates the absorptive part of these diagrams, then the on-shell
 condition for the
cut fermion line forces one of the exchanged gluons to have zero longitudinal
 momentum.
 Furthermore,
an explicit calculation \cite{BW} shows that the amplitude is indeed 
dominated
 by the
kinematic regime where one of the gluons is much more energetic than 
the other.
 As we shall see later,
an analogous infrared singularity also dominates the contribution of 
the graphs
 of
Fig. \ref{fig7}(a) to colour-singlet jet production in $e^+ e^-$ annihilation.
 Moreover, in
the case where one of the gluons in Fig. \ref{fig5} is soft, 
the quark-antiquark
 system
appears as a colour dipole, and the contributions from 
Figs. \ref{fig5}(a) and
 (b) cancel
each other if the quark and antiquark have large transverse momentum (and
 therefore
occupy a small region of impact-parameter space). Thus the process is 
dominated
 by the
low-transverse-momentum kinematic region in which  the fermions are almost
 parallel to
the incoming photon or hard gluon (this is the aligned jet model \cite{AJM}.
In this kinematic regime the internal quark is close
to its mass shell, so that the graph cannot be determined purely from
 perturbation theory.
An infrared scale, presumably  of order $\Lambda_{QCD}$, is introduced by the
 non-perturbative dynamics, and it is because of
this scale that the probability for large-rapidity-gap events is
not exponentially suppressed with rapidity gap, as would be expected from a
simple
dimensional analysis as in naive perturbation theory. Of course, since we are
looking at events at low $x$, the two gluons of Fig. \ref{fig5} constitute 
the
 Born term
of the complete BFKL  ladder, giving rise to  a ``hard'' Pomeron
with an intercept considerably above unity.
  Nevertheless, we expect that the above discussion will
hold when the complete ladder is taken into account.
The non-perturbative mechanism that gives
 rise to this infrared scale that prevents the large-rapidity-gap
 events from being exponentially suppressed, is the same mechanism
 that reduces the intercept of the ``soft'' Pomeron, which controls
 hadron diffractive processes at low momentum transfer, to a value close to
 unity.

In this paper we show that the  glueball production diagrams of Fig.
 \ref{fig7}(a), when
suitably regulated in the infrared region, do {\it not} give a large
 cross-section
for large-rapidity-gap events in $e^+ e^-$ annihilation.
Specifically, in the limit
$m/\sqrt{s}$ fixed, $s \rightarrow \infty$,
we find a cross section that falls as $1/s$, relative
to the normal three-jet cross section. This is the scaling law that one would
 normally expect
from dimensional analysis for the production of an isolated fixed-mass 
cluster,
 but
corresponds to a suppression of large-rapidity-gap events in $e^+ e^-$
 annihilation.
The reason for this difference from deep-inelastic-scattering
is that, after the regularisation of the
(logarithmic) infrared divergence by some suitable wavefunction or 
fragmentation
 function, the
events under consideration {\it can} be determined in perturbation theory, 
and
 the
infrared scale which is so important in the case of deep-inelastic scattering
 does not play
a significant role in the $e^+ e^-$-annihilation case, even though the 
relevant
 diagrams
can be thought of as the production of a ``Pomeron'' in the s-channel.

The layout of this paper is as follows.
In Section 2 we review the different kinematics of deep-inelastic scattering 
and
 $e^+ e^-$ annihilation,
discuss the evaluation of our basic perturbative QCD diagrams in Fig.
 \ref{fig7}, and
relate our calculation to that of ref.\cite{BBL}. Next, in Section 3, we 
discuss
 the treatment
of the infrared divergence in the cross-section, comparing and contrasting 
the
 $e^+ e^-$
situation with the use of naive Pomeron structure functions in deep-inelastic
 scattering,
and introducing an infrared cutoff derived from a physical picture of a
typical glueball wavefunction. Our main results
 are presented in section 4, including the dependences of the isolated 
 cluster
 cross section
on this wavefunction cutoff, on the cluster mass, on the rapidity gap 
with the
 rest of the
 event, and on the centre-of-mass energy, $\sqrt{s}$. Section 5 includes a
 discussion of
our results and the prospect for detecting large-rapidity-gap events in $e^+
 e^-$
annihilation. An Appendix presents the differential cross section for the
 production of two
quarks and two antiquarks in a suitable form for our study.

\section{The Parton-Level Calculation}
In $e^+ e^-$ annihilation, the equivalent of the ``pomeron'' exchanged in
the $t$-channel in a deep-inelastic scattering event is the production
 of a  colour-singlet gluon
cluster or a glueball in the case where the cluster consists of just one
 particle.
One of the Feynman diagrams for this process in leading order in perturbation
theory is shown in Fig. \ref{fig7}(a), in which the two gluons are 
constrained
to be in a colour singlet and to have fixed invariant mass, $m$, which
is much smaller than the total centre-of-mass energy $\sqrt{s}$.

A similar process was considered in ref.\cite{BBL}. However,
there the authors
constrained both the pairs of particles to have
 small invariant mass, thereby greatly reducing
the available phase-space and consequently reducing the cross-section.
In our case we will be concerned with the differential cross section with
 respect
to $m^2$, subject to the requirement that there must be a minimum
rapidity gap, $\Delta y$ between the gluon cluster and either of the fermion
jets.
\beq \Delta y = \mbox{ min}(\Delta y_1, \Delta y_2) \label{eq21} \eeq
If the quark (antiquark) jet $i, \ (i=1,2)$ makes an angle $\theta_i$
with the direction of the gluon cluster, then the rapidity gap
between that jet and the gluon cluster is (up to corrections of
order $m^2/s$)
\beq \Delta y_i = \ln \left( \frac{\sqrt{s}}{m} \right) + \ln x_3
   - \frac{1}{2} \ln \left( \frac{1+\cos \theta_i}{1-\cos \theta_i} \right)
\label{eq22}  \eeq
where the variable
$x_3$ introduced in ref. \cite{DEFG} is the fraction of available energy
 ($\sqrt{s}/2$) of the
gluon cluster. We may rewrite this in terms of the  energy fraction,
$x_i$ carried by the $i^{th}$ fermion jet, using the relation
$$ (1-\cos \theta_i) =  \frac{2 (x_i+x_3-1)}{x_i x_3} $$

\noindent For very large rapidity gaps, the fermion jets are both forced to 
be
 almost in the opposite direction to the gluon cluster and in this
limit we recover the results of ref. \cite{BBL}.

In $e^+ e^-$-annihilation,
the kinematic quantity which is analogous to the
invariant mass of the quark-antiquark pair produced in the
photon-pomeron collision is given by

\beq M_X^2 =  s(1-x_3) \eeq
The fact that we wish to allow this to be of order $\sqrt{s}$ means that
we stay away from the $x_3 \rightarrow 1$ limit. On the other hand,
as can be seen from Eq.(\ref{eq22}),
requiring a very large rapidity gap
forces $x_3$ to be close to unity. As we shall see in Section 4, the 
production
rate drops rapidly in this limit.

In more detail, if we assign momenta $p_1$ and $p_2$ to the outgoing quark
and antiquark  and momenta $p_3, \ p_4$ to the two gluons, then we
fix
\beq s_{34}=m^2 \eeq
and identify
\beq s_{134}=s(1-x_2) \eeq
and
\beq  s_{234}=s(1-x_1) \eeq
where
$$ s_{ij}=(p_i+p_j)^2 $$
and
$$ s_{ijk}=(p_i+p_j+p_k)^2$$
This enables us to select the required region of phase space.
One further quantity that we shall need to identify is $z$, the fraction
of the energy of the gluon cluster carried by one of the gluons.
Up to corrections of order $m^2/s$, this is given by
\beq z \approx \frac{s_{13}}{s_{13}+s_{14}}
  \approx \frac{s_{23}}{s_{23}+s_{24}} \label{zeq} \eeq

The squared matrix element for the processes shown in Figs. \ref{fig7} (a)
and (b) are given in the Appendix of ref.\cite{ERT}. What we need to do is
to identify the relevant colour factors for the process we are considering,
and perform a (numerical)  integral over the required region of phase
space. In ref.\cite{ERT} a summation was performed over all final-state
helicities, in particular over the final-state helicities of the gluons.
If the gluon cluster consists of a single glueball, we might wish
to project out a particular linear combination of helicity states
which make up the spin of the glueball. In the absence of any concrete
information about what spin states would be expected to dominate
for particular gluon cluster masses we do not take this into
consideration, and we assume that summing over all gluon helicity
states does not introduce significant errors.

In order to project out the colour-singlet part of the gluons
(of colours $a$ and $b$) we apply
$$ \frac{\delta_{ab}}{\sqrt{8}} $$
to the matrix element. Thus the colour factor for the squared matrix element
 becomes
\beq (\tau^a \tau^b)_{ij} \frac{\delta_{ab}}{\sqrt{8}}
 (\tau^c \tau^d)_{ij} \frac{\delta_{cd}}{\sqrt{8}} =
  \frac{C_F^2}{8} \delta_{ij} \delta_{ij} \eeq
where the $\tau^a$ are the colour matrices and $C_F=4/3$
is the quadratic Casimir
in the fundamental representation. The factor $\delta_{ij} \delta_{ij}$
is present in the tree-level total cross section to which we
normalise our calculations. Thus the rule is that we set $C_A=0$ in the
formulae given in \cite{ERT} and divide by a factor of 8.

For the process shown in Fig. \ref{fig7}(b) the colour factor is identical.
The difference is that we now set
\beq  s_{12}=m^2 \eeq
and make similar changes ($1 \leftrightarrow 3, \ 2 \leftrightarrow 4$)
in the rest of the kinematics.

For the process shown in Fig. \ref{fig7}(c) the complete squared matrix 
element
 was not given in \cite{ERT}, since some of the interference
terms vanish when integrated over phase-space by virtue of Furry's
theorem. However, this cancellation only occurs if the phase-space
integration is performed to find the differential cross-section
with respect to a variable which is symmetric in all the final
state particles. This is not the case in the process we are considering,
since we require that two of the fermions should have a small invariant mass.
The squared matrix element  for the production of two quark-antiquark
colour signlet pairs each with flavours a and b
(including the colour factor) is given in the Appendix. It is
necessary to perform a sum over all possible flavours and normalise to
the total hadronic cross section as discussed in ref.\cite{BBL}.

Requiring a large rapidity gap eliminates the region of phase space
where one finds collinear divergences, in which a gluon runs
parallel to a fermion line. However, we do not eliminate infrared divergences
which occur when one of the gluons becomes soft, i.e. when the fraction
of energy, $z$, of the gluon cluster,  carried by one of the
gluons goes to zero or unity. This introduces a logarithmic divergence
in the phase-space integral. Such divergences are regularized by the
fragmentation of the two gluons into hadrons, or by a wavefunction
in the case of a single glueball  state. The fragmentation function
or wavefunction must vanish when $z=0$ or $z=1$. It is to a discussion
of this wavefunction that we now turn.

\section{The Wavefunction}

One example of a possible form for the two-gluon squared wave function
of a glueball
would be
\beq
|\psi(z)^2| = {\cal N} z(1-z)
\eeq
where ${\cal N}$ is a normalization factor and $z$ and $(1-z)$ are the
longitudinal
momentum fractions of the constituent gluons in the infinite-momentum
frame.  A
similar form has often been discussed in connection with the
deep-inelastic Pomeron
structure function, and it has the desirable feature of removing the
infrared
singularity as $z \rightarrow 0,1$.  However, the physical intuition
behind this choice
is not very clear, and we prefer to use a wave function that is better
motivated by
simple ideas about the physical composition of a glueball.

Two gluons with negligible mass, equal and opposite transverse momenta
$\underline{k}_T$, and longitudinal momentum fractions $z$ and $(1-z)$ in the
infinite-momentum frame, have a combined invariant mass-squared
\beq
m^2 = \underline{k}^2_T \left( \frac{1}{z} + \frac{1}{1-z} \right)
\eeq
The corresponding transverse size $R_T$ is given by
\beq
R^2_T = \frac{1}{\underline{k}_T} = \frac{1}{z(1-z)}~\frac{1}{m^2}
\eeq
We choose the following plausible form for the glueball wave
function
\beq
|\psi(R^2_T)| \approx {\rm exp} \left( \frac{-R^2_T}{b^2} \right)
\eeq
for some size parameter $b$.  The corresponding squared
longitudinal-momentum wave
function is
\beq
|\psi(z)^2| = {\cal N} {\rm exp} \left[ - \frac{1}{m^2b^2z(1-z)} \right]
\label{wvfn} \eeq
where ${\cal N}$ is a normalization factor.
We note that this also regulates the infrared divergences from the
parton-level differential cross-section when $z \rightarrow 0,1$.
Possible choices of $b$ for
different
glueball masses $m$ will be discussed in the next section.

Clearly, other choices of wave function are possible, which injects an
uncontrolled
infrared sensitivity into our results.  Nevertheless, we believe that
the results we
present in later sections reflect qualitatively what might be expected
with any
``reasonable" choice of wave function.


\section{Results}

In this section we present our main results.

\begin{figure}
\begin{center}
\leavevmode
\hbox{\epsfxsize=3.5 in
        \epsfysize=2.9 in
\epsfbox{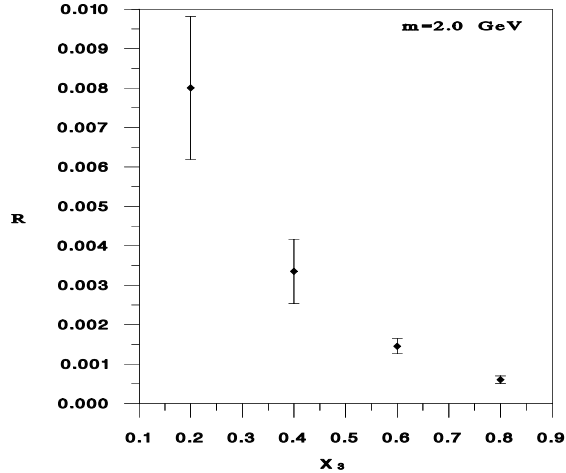}}
\end{center}
\caption{Graph of $R$ defined in Eq.(4.1) as a function of
$x_3$. The error
bars show the variation as $x_1$ (or $x_2$) is varied keeping
$x_3$ fixed.}  \label{fig8} \end{figure}
We begin by demonstrating that the cross section for the production
of a low-mass gluonic cluster and two fermion jets depends mainly
on the energy fraction of the gluon cluster  $x_3$ and rather less
on $x_1 \ (x_2)$, \footnote{We remind the reader that $x_1,x_2,x_3$
are related by $x_1+x_2+x_3=2$}
 the energy fractions of the fermion jets. We show
this in Fig. \ref{fig8}, in which the quantity $R$ is the ratio
\beq R= \frac{d^3 \sigma}{dx_1 dx_2 dm^2} / \frac{d^2 \sigma_0}{dx_1 dx_2}
 \label{eq41}\eeq
where $d^2 \sigma_0/dx_1 dx_2$ is the tree level cross section
for the production of three-jets calculated in ref. \cite{DEFG}. We have 
taken
the gluon cluster mass, $m$, to be $2 \ GeV$ and the centre-of-mass energy 
equal
 to
$M_Z$. The infrared divergence which occurs when one of the gluons
becomes soft is regulated simply by demanding that all invariant masses
should be greater or equal to $m$. The graph clearly shows a substantial
dependence on $x_3$ and a rather modest dependence on $x_1 \  (x_2)$.
Since the production rate is dominated by the region of phase space where one
of the gluons is soft, this is at first a rather surprising result, since
we would expect that an extra soft gluon would yield an $x_3$ dependence 
which
was not very different from the three-jet $x_3$ dependence, so that
one  might have expected
that this ratio should be approximately constant. However,
as explained above, this is {\it not} the case for deep-inelastic scattering,
in which the ratio of the two-gluon to one-gluon exchange is very sensitive
to the transverse momentum of the fermion pair. The $x_3$ dependence that
we see here is the analogue of this effect in $e^+ e^-$ annihilation.

We now turn to the more realistic case of the perturbative prediction
of events with a low-mass gluonic cluster in which the infrared divergence
is regulated by a wavefunction as discussed in Section 3. We take as
the default value of the ``average impact parameter''
$b=2 \ GeV^{-1}$, and again the cluster mass is $2 \ GeV$. We start
by showing distributions not with a minimum rapidity gap, but with a minimum
opening angle $\theta$ between the gluon cluster and either of the fermion
jets. Such opening angles are more directly controlled experimentally.
These opening angles are related to  the rapidity gap, but the
exact relation between the opening angle and the rapidity gap
depends on the gluon cluster energy fraction $x_3$.

\begin{figure}
\begin{center}
\leavevmode
\hbox{\epsfxsize=3.7 in \epsfysize=3.0 in
\epsfbox{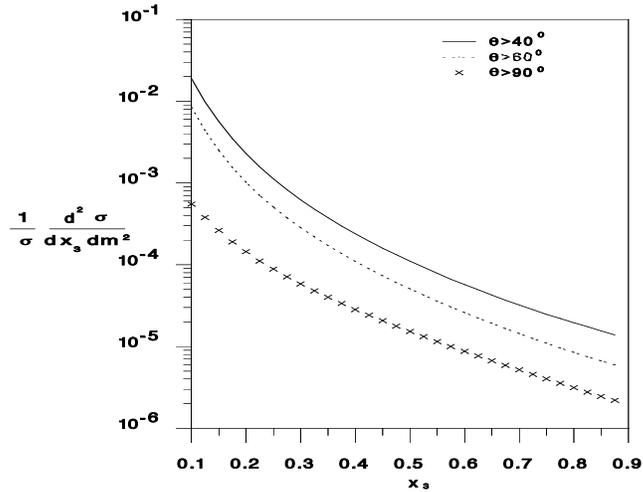}}
\end{center}
\caption{Graph of differential cross section  against $x_3$
for different minimum opening angles between the gluon cluster and the
fermion jets.}  \label{fig9} \end{figure}

\begin{figure}
\begin{center}
\leavevmode
\hbox{\epsfxsize=3.7 in \epsfysize=3.0 in
\epsfbox{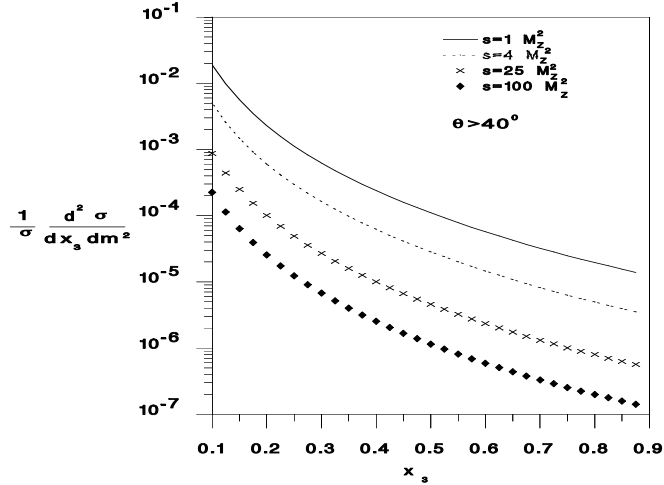}}
\end{center}
\caption{Graph of differential cross section  against $x_3$
for different centre-of-mass energies.}  \label{fig10} \end{figure}

The results are shown in Fig. \ref{fig9}, where we have taken three different
minimum opening angles. The substantial decrease in differential cross 
section
as the opening angle is increased is indicative of a substantial decrease
with increasing rapidity gap.
We emphasize this point further in Fig. \ref{fig10}, where we fix the
minimum opening angle to be $40^0$, but allow the centre-of-mass energy

to take values $2 \ M_Z$, $5 \ M_Z$ and $10 \ M_Z$ as well as $M_Z$.
The sharp decrease in differential cross section goes like $1/s$, as
expected from dimensional analysis, which means that
events with a low-mass gluon cluster will be even more difficult to detect
in higher-energy $e^+ e^-$ machines in the future.

\begin{figure}
\begin{center}
\leavevmode
\hbox{\epsfxsize=3.7 in \epsfysize=3.0 in
\epsfbox{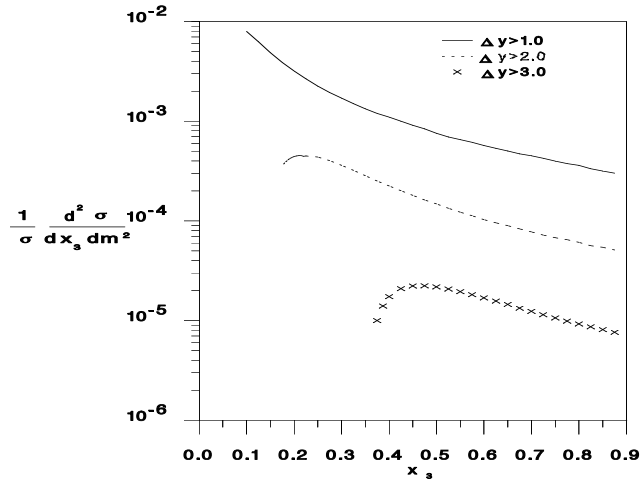}}
\end{center}
\caption{Graph of differential cross section  against $x_3$
for different minimum rapidity gaps.}  \label{fig11} \end{figure}

\begin{figure}
\begin{center}
\leavevmode
\hbox{\epsfxsize=3.7 in \epsfysize=3.0 in
\epsfbox{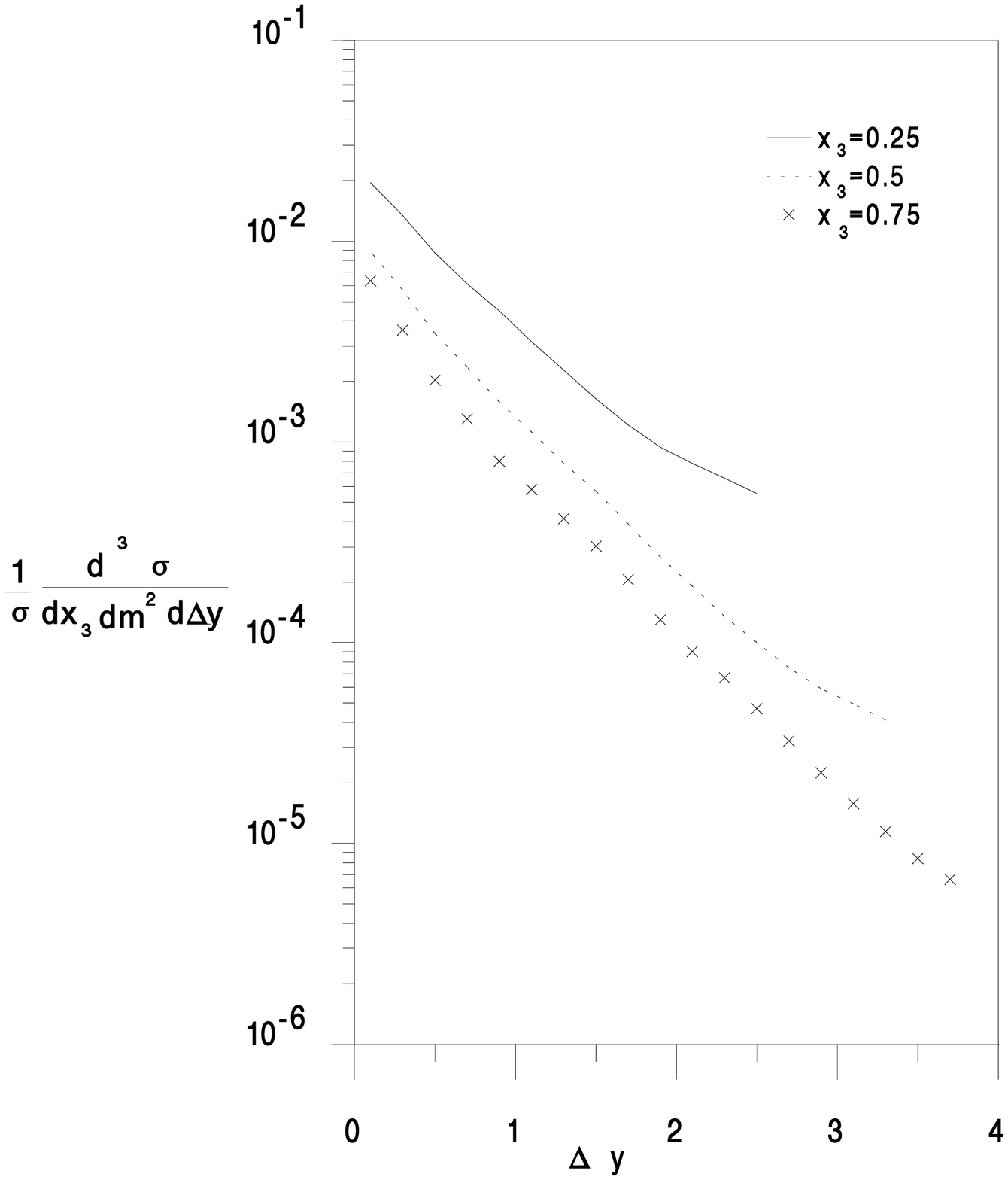}}
\end{center}
\caption{Graph of differential cross section  against minimum
 rapidity gap for various different values of $x_3$.}
 \label{fig12} \end{figure}

Next we turn to  Fig. \ref{fig11}, where
we plot the differential cross-section, again against $x_3$, for different
values of the minimum rapidity gap.
A behaviour  $\sim \exp  (-c \Delta y)$, with $c \approx 2$ can be seen.
Once again, this is the  $1/s$ behaviour
that one would expect from
 dimensional analysis,  which is therefore expected to hold
(up to logarithmic corrections) in any perturbative calculation.
In Fig. \ref{fig12} we show the plot the other way around, i.e.,
for fixed gluon cluster energy fraction $x_3$ we plot the
differential cross-section with respect to $m^2$ and $\Delta y$.
The same fall-off with the length of the rapidity gap
can be seen.  It is worth noting
in both of these graphs that there is a sharp cutoff in $x_3$
for a given rapidity gap. This is because
                              events with the required
rapidity gap are excluded by the kinematics if the energy of the
gluon cluster is below a certain value.

\begin{figure}
\begin{center}
\leavevmode
\hbox{\epsfxsize=3.7 in  \epsfysize=3.0 in
\epsfbox{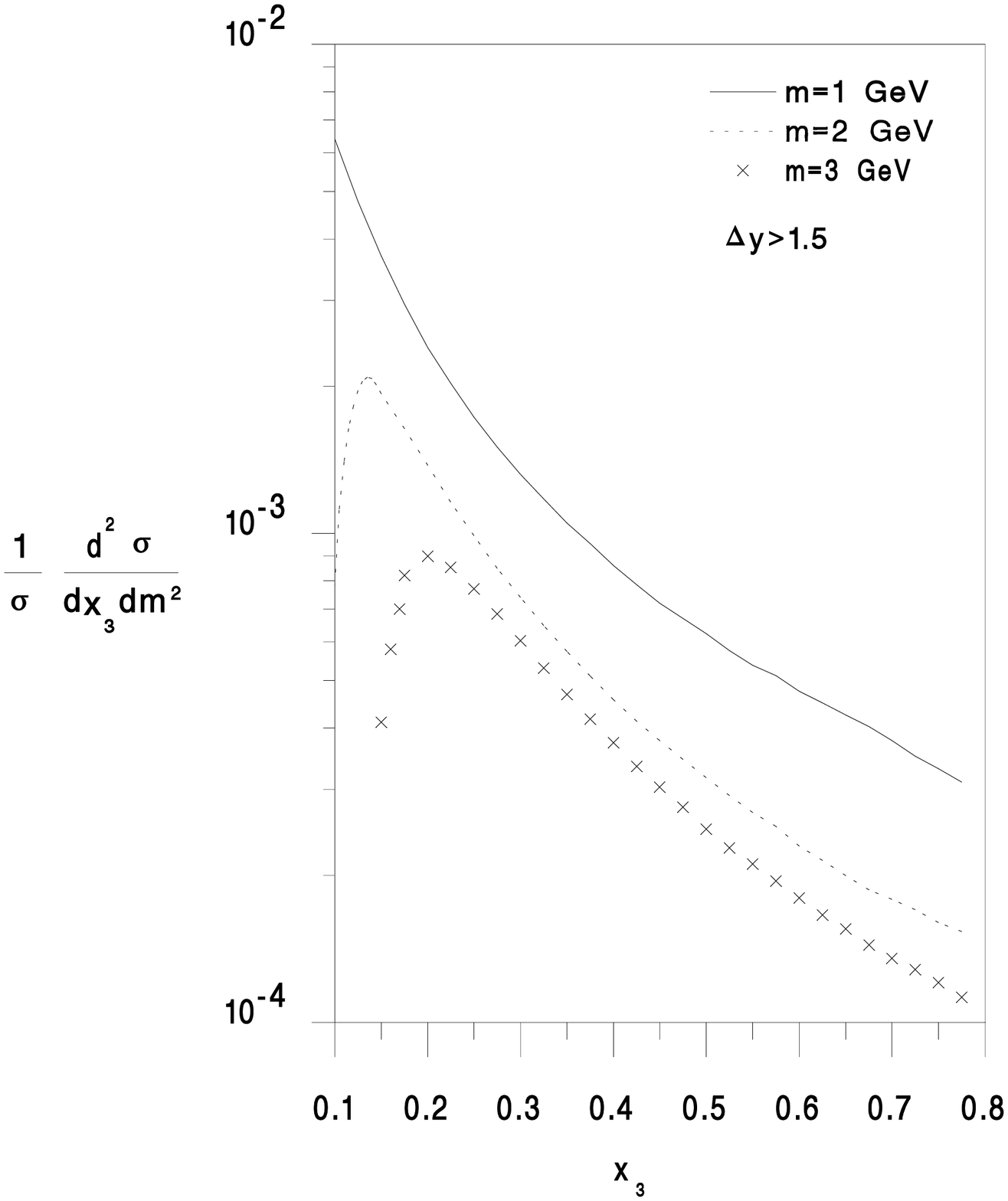}}
\end{center}
\caption{Graph of differential cross section  $x_3$
for different values of $m$.}
 \label{fig13} \end{figure}

\begin{figure}
\begin{center}
\leavevmode
\hbox{\epsfxsize=3.7 in \epsfysize=3.0 in
\epsfbox{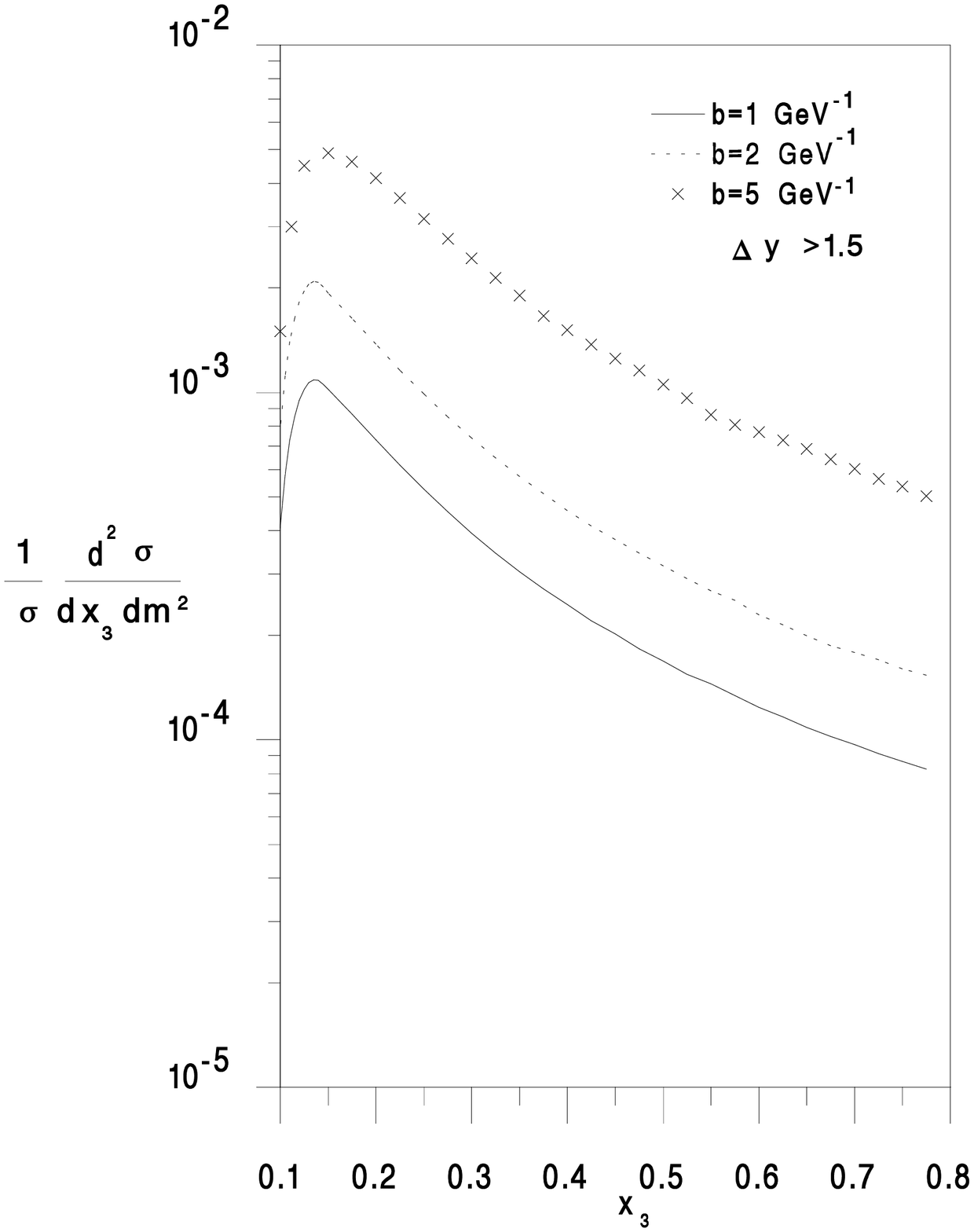}}
\end{center}
\caption{Graph of differential cross section   $x_3$
for different values of $b$.}
 \label{fig14} \end{figure}

In order to demonstrate the dependence on the other two parameters,
namely the mass $m$ of the gluonic cluster and the
``average impact parameter'' $b$
used in the wavefunction, we plot in Figs. \ref{fig13} and \ref{fig14}
the differential cross-section with the minimum rapidity gap (set to
$\Delta y=1.5$) for different values of $m$ and $b$ respectively.
As expected, if we increase the cluster mass $m$  there is
more  phase space available for the events, and so the
differential cross-section is increased. As $b$ is increased,
the wavefunction (see Eq.(\ref{wvfn})) provides
a suppression only for fractional gluon energy $z$
closer to the end-points $z=0$ and $z=1$. We therefore pick up
more of the infrared-enhanced differential cross-section and this
explains the increase seen in Fig.\ref{fig14}.

\begin{figure}
\begin{center}
\leavevmode
\hbox{\epsfxsize=3.7 in \epsfysize=3.0 in
\epsfbox{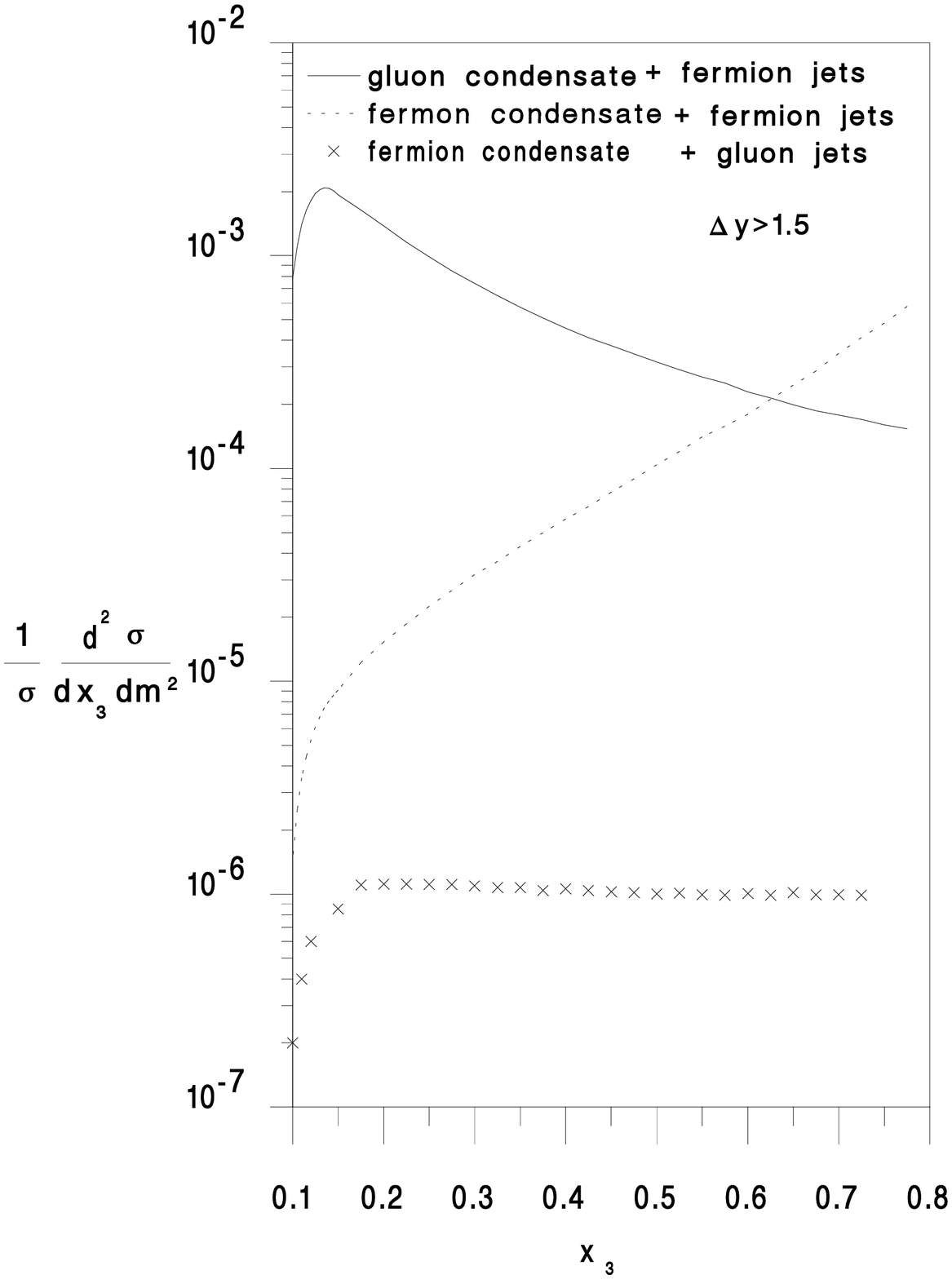}}
\end{center}
\caption{Graph of differential cross section   $x_3$
showing gluon-cluster and fermion-cluster contributions.}
 \label{fig15} \end{figure}

Finally, we look at the contributions to a colour-singlet cluster
formation from quark-antiquark pairs. There are two possibilities
shown in Figs. \ref{fig7}(b) and (c). In one case the other two jets
are gluonic - here we consider the same matrix element given in
the Appendix of ref. \cite{ERT}, but in a different region of phase
space. In the other case the other jets are also fermionic. The
squared matrix element  for this process is given in the Appendix.
The results are shown in Fig. \ref{fig15}, in which we
have taken the default values, $m=2 \ GeV, \ b=2 \ GeV^{-1}$ and
a rapidity-gap of 1.5. We see that, apart from large $x_3$ where the
gluon cluster contribution is small, both the  fermion
cluster contributions are negligible and do not show the same
strong dependence on $x_3$ as the gluon cluster. The reason for
this is that there is no infrared-enhanced contribution.
In the first case, where the unobserved jets are gluonic,
since the quark and antiquark are almost parallel the soft
gluon insertion cancels between the insertion on the
quark and antiquark line (in a similar way to the
cancellation at large $p_\perp$ in deep-inelastic scattering).
In the case of a four-fermion final state, there is no
infrared divergence when we force the quark and antiquark with the
same flavours to go into different colour-singlet clusters.

In general, the differential cross-sections we have found are
very small and  cannot be observed at present. However, they are
considerably larger than those predicted in ref. \cite{BBL}, which
is hardly surprising since in our case there is much
more  phase space available.

 \section{Discussion}
We have considered in this paper the possible rate of large-rapidity-gap
events in
$e^+e^-$ annihilation due to the lowest-order perturbative QCD diagrams
shown in
Fig.~7.  The kinematical configuration we study (Fig.~3) is similar to
that of the
rapidity-gap events observed at HERA (Fig.~1), and more general than the
case of two
low-mass clusters considered previously (Fig.~2).  Correspondingly, we
find a
cross-section that is considerably larger than that estimated in Ref.
\cite{BBL}.  Taking
the ratio $R$ of colour-singlet cluster events to conventional three-jet
events
(Fig.~6), we find that $R$ depends sensitively on the scaled gluon energy
$x_3$, but
less sensitively on the scaled $q$ and $\bar q$ energies $x_1, x_2$
(Fig. 8).  As we
show in Figs.~9, 11 and 12, the cross-section falls as the angle
(rapidity) gap
between the colour-singlet cluster and the rest of the event is
increased:  there is
no indication of a ``rapidity plateau" as known in hadron-hadron and
$ep$ collisions.
Correspondingly, we see in Fig. 10 that the cross section falls as the
centre-of-mass
energy increases.  This is not surprising, since an infrared cutoff
scale appears
via the fixed mass $m$ of the colour-singlet cluster
considered:  the $m$-dependence of
the cross section is shown in Fig.~13.  As seen in Fig.~15, most of the
colour-singlet clusters are digluons.
Our cross-section estimates are
based upon a
particular form for the infrared cutoff discussed in Section 3:  the
sensitivity to
the glueball wave-function size parameter $b$ is illustrated in Fig.~14.
However,
other choices of the form of infra-red cut-off are possible, and the
uncertainty in
the cross-section estimates could well be larger than suggested by
Fig.~14.
Nevertheless, we find that the cross section for large-rapidity-gap
events probably
lies below the experimental sensitivity of, for example, a recent search
by the ALEPH
collaboration \cite{Schmitt}. The fact that the
calculated
cross section falls with the centre-of-mass energy, as seen in Fig.~10,
means that
this situation will worsen at higher energies (e.g., LEP II,
NLC/JLC/CLIC), despite
the larger phase space.

We therefore conclude that the prospects for observing
large-rapidity-gap events in
$e^+e^-$ annihilation are dim, at least in the context of perturbative
QCD.  This
reflects the different dynamical conditions in the time-like region,
namely
colour-singlet production $e^+e^-$ annihilation, and the space-like
region, namely
crossed-channel colour-singlet exchange in hadron-hadron or $ep$
collisions.  One has
to be ``lucky" to produce two real gluons close together in phase space,
with none
radiated elsewhere, whereas it is relatively ``easy" for a long-range
Coulomb gluon
field to bleach colour over a large rapidity gap.     \bigskip

 \section{Acknowledgements:}
 The authors are grateful to Michael Schmitt for useful conversations.

 One of us (J.E.) would like to thank the Physics Department at Southampton
 University for its hospitality   while part of this work was done.

\newpage
\section{Appendix}
\setcounter{equation}{0}
\renewcommand{\theequation}{A.\arabic{equation}}
In this Appendix we present the tree-level result for the differential cross
section $d\sigma$ for an $e^+ e^-$ pair with centre-of-mass energy
$\sqrt{s}$ to go into a colour-singlet pair consisting of a quark with 
momentun
 $p_1$
and flavour $a$ (charge $Q_a$)
plus an antiquark with momentum $p_2$ and flavour $b$
(charge $Q_b$), and a colour singlet
consisting of a quark with momentum $p_3$ and flavour $b$ plus an antiquark 
with
 momentum
$p_4$ and flavour $a$.

Defining
$$ s_{ij}=(p_i+p_j)^2 $$ $$s_{ijk}=(p_i+p_j+p_k)^2 $$
we have (normalising with respect to the leading-order total cross-section,
 $\sigma_0$)
\begin{eqnarray}
\frac{1}{\sigma_0} d\sigma &=&   \frac{1}{\sum_a Q_a^2}
 \frac{C_F^2}{3} \left( \frac{\alpha_s}{2\pi} \right)^2
   \frac{1}{4s^2} \int ds_{134} \int ds_{234} \int ds_{34}   \theta (s_{134}
 s_{234}- s s_{34})
\nonumber \\ & &   \theta (s_{34}+s-s_{134}-s_{234} ) \int_0^1 dv
       \frac{1}{\pi} \int_0^\pi d\phi \left[
        f(s_{12},s_{13},s_{14},s_{23},s_{24},s_{34})+ \right.
         \nonumber \\ & & \left.
             f(s_{34},s_{24},s_{14},s_{23},s_{13},s_{12})+
     f(s_{34},s_{13},s_{23},s_{14},s_{24},s_{12})+ \right. \nonumber \\ & &
 \left.
          f(s_{12},s_{24},s_{23},s_{14},s_{13},s_{34})   \right] \end{eqnarray}
where
$$ v= \frac{s_{24}}{s_{124}-s_{34}}, $$
 $\phi$ is the azimuthal angle of the $34$ system relative to the $12$ plane,
 and
\begin{eqnarray}
      f(s_{12},s_{13},s_{14},s_{23},s_{24},s_{34})  
      &=&(Q_a^2 + Q_b^2) \left[
    \frac{1}{s_{14}^2 s_{134}^2}
    \left(  s_{12} s_{13} s_{14} + s_{12} s_{13} s_{34}
        \right. \right.
         \nonumber \\ & & \hspace*{-125 pt}
      \left. \left.    -s_{12} s_{34}^2
      - s_{13}^2 s_{24} + s_{13} s_{14} s_{23}
   + 2 s_{13} s_{23} s_{34} + s_{13} s_{24} s_{34} + s_{14} s_{23} s_{34}
           \right. \right. \nonumber \\ & & \hspace*{-100pt} \left. \left.
              + s_{14} s_{24} s_{34} \right)
          \right. \nonumber \\ & &  \hspace*{-125 pt} \left.
         +     \frac{1}{2 s_{14}^2 s_{134} s_{234}} \left(
       - s_{12}^2 s_{34} + s_{12} s_{13} s_{24} + s_{12} s_{13} s_{34}
       + s_{12} s_{14} s_{23} - 2 s_{12} s_{23} s_{34}
     \right. \right. \nonumber \\ & &  \hspace*{-130 pt} \left. \left.
        +s_{12} s_{24} s_{34}
        - s_{12} s_{34}^2 - s_{13}^2 s_{24} + s_{13} s_{14} s_{23}
         - 2 s_{13} s_{23} s_{24} -  s_{13} s_{24}^2
   + s_{13} s_{24} s_{34}
          \right. \right.  \nonumber \\ & & \hspace*{-125 pt} \left. \left.
       +2 s_{14}^2 s_{23}
       + 2 s_{14} s_{23}^2 + s_{14} s_{23} s_{24} + s_{14} s_{23} s_{34} 
       \right)
        \right] \nonumber \\ & &
       \hspace*{-1.5in } -  Q_a Q_b \frac{1}{4 s_{14} s_{23}}
     \left[    \frac{1}{s_{124} s_{123}}  \left(
       -  s_{12}^2 s_{34} - 4 s_{12} s_{13} s_{14} -
         5 s_{12} s_{13} s_{24}
           \right. \right. \nonumber \\ & & \hspace*{-125pt} \left. \left.
       -3 s_{12} s_{14} s_{23} -
    2 s_{12} s_{14} s_{34} - 4 s_{12} s_{23} s_{24}
       - 2 s_{12} s_{23} s_{34} - 4 s_{12} s_{24} s_{34}
         \right. \right. \nonumber \\ & & \hspace*{-125pt} \left. \left.
         -2 s_{12} s_{34}^2
         -2 s_{13} s_{14} s_{24}
      - 2 s_{13} s_{23} s_{24} - s_{13} s_{24}^2 + 2 s_{13} s_{24}
      s_{34}
        + 2 s_{14}^2 s_{23}
       \right. \right.  \nonumber \\ & & \hspace*{-125 pt} \left. \left.
           +2 s_{14} s_{23}^2 +s_{14} s_{23} s_{24}
      + 2 s_{14} s_{23} s_{34}
          \right)
           \right. \nonumber \\ & & \hspace*{-100pt} \left.
        + \frac{1}{s_{124} s_{234}} \left( -2 s_{12} s_{13} s_{34}
        + 2 s_{12} s_{14} s_{34} + 4 s_{12} s_{23} s_{24} +
        2 s_{12} s_{23} s_{34}
     \right. \right. \nonumber \\ & & \hspace*{-125pt} \left. \left.
      + 6 s_{12} s_{24} s_{34}
      +2 s_{13} s_{24}
      - 2 s_{13} s_{14} s_{23} + 2 s_{13} s_{14} s_{24}
            + 2 s_{13} s_{23} s_{24}
      \right. \right. \nonumber \\ & & \hspace*{-125pt} \left. \left.
           +2 s_{13}s_{24} -2 s_{14}^2 s_{23}
            -2 s_{14} s_{23}^2
            + 2 s_{14} s_{23} s_{24} + 4 s_{14} s_{24} s_{34}
            \right) \right]  \end{eqnarray}
The terms proportional to $(Q_a^2+Q_b^2)$ coincide with
the quantity $D$ of \cite{ERT} (Eq.B.6), with the colour factor $T_R$ 
replaced
 by $C_F/3$.
The other terms are the class $F$ terms (see Table 2 of ref.\cite{ERT}),
which were not given explicitly in ref.\cite{ERT}.

\newpage

\end{document}